# Detection of per- and polyfluoroalkyl water contaminants with multiplexed 4D microcavities sensor

ANTON V. SAETCHNIKOV[1,*], ELINA A. TCHERNIAVSKAIA[2], VLADIMIR A. SAETCHNIKOV[3], AND ANDREAS OSTENDORF[1]

[1]*Chair of Applied Laser Technologies, Ruhr University Bochum, 44801 Bochum, Germany*
[2]*Physics Department, Belarusian State University, 220030 Minsk, Belarus*
[3]*Radio Physics Department, Belarusian State University, 220064 Minsk, Belarus*
[*]*anton.saetchnikov@rub.de*

**The per- and polyfluoroalkyl substances (PFAS) constitute a group of organofluorine chemicals treated as the emerging pollutants and currently are of particularly acute concern. These compounds have been employed intensively as surfactants over multiple decades and are already to be found in surface and ground waters at amounts sufficient to have an effect on the human health and ecosystems. Because of the carbon-fluorine bonds the PFAS have an extreme environmental persistence and their negative impact accumulates with further production and penetration into the environment. In Germany alone, more than thousands sites have been identified to be contaminated with PFAS and thus timely detection of PFAS residues is becoming a high-priority task. In this paper we report on the high performance optical detection method based on whispering gallery modes microcavities applied for the first time for detection of the PFAS contaminants in aqueous solutions. A self-sensing boosted 4D microcavity fabricated with two-photon polymerization is employed as an individual sensing unit. On example of the multiplexed imaging sensor with multiple hundreds of simultaneously interrogated microcavities we demonstrate the possibility to detect the PFAS chemicals representatives at the level of down to 1 ppb.**

## Introduction

A broad family of organofluorine contaminants known as per- and polyfluoroalkyl substances (PFAS) is a group of more than 10 thousands of engineered chemicals. They consist of a fully or partially fluorinated carbon chain of various length with different functional groups (predominantly carboxylate or sulfonate). These substances are water-, grease- and dirt-repellent agents that have been vastly used during the last four decades as surfactants in numerous consumer products. Existence of the fluoroalkyl tail, strong carbon-fluorine bond, and exceptionally high thermal and chemical stability do not allow PFAS to degrade via water, light, or bacteria. For this reason, PFAS have been classified as persistent organic pollutants and the more they are produced and enter the environment, the more they are accumulated. PFAS have already penetrated into the soil, atmospehere, and groundwater, and thus became hazardous for the ecosystems, biodiversity, and human health [22, 12]. Particuraly, numerous studies show potential immunotoxicity of the PFAS chemicals, their carcinogenicity, effect on fertility, and interaction with the hormonal system [2]. The above-mentioned has prioritized PFAS chemicals for complete elimination.

The traditional analytical approaches such as chromatography (HPLC) or gas (GS) coupled with mass spectrometry (MS) are the most common methods for PFAS detection at concentrations down to ng/l [34, 27]. Despite the excellent performance in terms of accuracy, precision, intrinsic multiplexity, and specificity, they suffer from apparatus cost and complexity, require laboratory conditions and skilled personnel, and imply time-consuming processes for sample preparation and data collection. To address these drawbacks, a quick, straightforward, and sensitive technique for PFAS chemical detection is demanded. An appropriate alternative to the conventional laboratory devices can be provided by the biochemical sensors, which are based on various physical principles and offer improved selectivity, flexibility, and compactness for the real-time online monitoring. Among them, the optical sensors with their high efficiency of light-matter interaction and electromagnetic interference immunity are worthy of special attention. Optical sensor-based solutions for PFAS detection reported so far are attributed to the fiber/waveguide-based configurations using sufrace plasmon resonance (SPR) [5, 6, 25], lossy mode resonance [21], and interferomentic schemes [11]. Particular solutions employ functional layers out of polyvinylidene fluoride [11], antibodies [6], serum albumin [21], or molecularly imprinted polymers [5, 25] for enhanced PFAS sensitivity.

Although the fiber/waveguide sensor-based configurations provide low detection limit (reported from ∼ 0.1 ppb to several ppm), their sensitivity remains restricted by the limited interaction length between the evanescent field of the guided mode and chemical compounds, where the major part of the evanescent tail remains unaffected and hence does not contribute to the signal. The sensing capability enhances when the fiber/waveguide-based detection is realized within an interferometric scheme to enable quantification of the optical field phase changes at the cost of the detection principle complexity. In the SPR-based approach the metal surfaces are strongly influenced by the temperature variations and due to high absorption, the surface plasmon wave attenuates rapidly in the propagation direction, limiting the interaction length even more than in fiber/waveguide-based configurations. Here the optical microresonator-based sensing with its optical field interaction length with the sensed solution higher by orders of magnitude represents a competitive alternative for the PFAS chemicals detection.

For optical microresonators the EM field is trapped by the contrast in refractive indices inside the dielectric material with closed circular loop [3, 38]. For those waves that travel along the cavity's edge and interfere with each other after a single roundtrip, the field accumulates in the cavity. Whispering gallery mode (WGM) resonances are the eigenmodes that satisfy the constructive interference condition. These resonances are characterized by the low roundtrip energy dissipation and correspondingly by high quality factors (Q-factor) [17]. Due to the presence of the evanescent field outside the cavity geometry, the WGM's spectral properties become susceptible to changes of morphology, material, and/or geometry of the resonator as well as variations in the environment [13, 4, 41].

The microresonator-based sensing technology underwent significant advancements over the last decades that enabled sensing of different physical and chemical parameters [24, 36, 10, 30, 18, 19], and with possibility for deep-learning powered multiplexed sensing of hundreds or even thousands of resonators [32, 28]. Since the first demonstration of biochemical sensing with WGM microresonators [39], multiple different method extensions have been proposed for further boosting of sensitivity. Among them are the plasmonic-photonic scheme [8, 35], activation (doping cavity with gain medium) [26, 37], exceptional points [7] and material response enhanced sensing by realization of 4D structures-based sensing [29]. The latter represents the self-sensing phenomenon that can be initiated under external stimuli on 3D microresonators manufactured with two-photon polymerization (2PP) [23]. As a result, the spectral response of the sensor becomes amplified by induced reversible variations of the resonator and can be tracked within WGM detection scheme at high precision. Nevertheless, despite the numerous benefits over other optical sensing techniques [40], the microresonator-based detection has not been employed for PFAS detection so far.

In this paper we report on detection of PFAS chemicals in water with the optical microresonators sensing approach demonstrated for the first time to the best of our knowledge. The WGM-based detection is realized in multiplexed imaging configuration with 4D microresonators as individual sensing units. To gain the high sensing performance, an approach of allocating the water-matched sub-wavelength film onto the substrate prior to two-photon polymerization is proposed and the impact of deposited layer onto the sensor fabrication process is numerically studied. We demonstrate that the pre-processed substrate enables enhancement of the loaded Q-factor of 2PP microresonator in the water up to $10^5$. Thanks to the self-sensing phenomenon of the resonator photoresin the spectral response of the multiplexed microcavity detector improves by orders of magnitude relative to the pure evanescent-field sensing. PFAS detection with 4D microresonators-based approach is validated on examples of two anionic contaminants with carboxylate or sulfonate polar groups and different chain length. We demonstrate that the response of the multiplexed 4D microresonator sensor on at least 10 ppb of perfluorooctanoic acid (PFOA) and 1 ppb of perfluorobutanesulfonic acid (PFBS) clearly stand out relative to the water without the use of any functional layers.

## Results and discussion

**4D microresonator model** The model of the individual microresonator for fabrication with two-photon polymerization is chosen similar as reported earlier in [29]. Its major and minor radii ($R$ = 21 $\mu$m and $\rho$ = 1.8 $\mu$m) are numerically optimized to minimize the radiation losses on the microcavity's curvature. With the selected geometry the radiation limited Q-factor approaches to the absorption limited one for SZ2080 polymer ($1.3 \times 10^7$) utilized for manufacturing. In order to minimize the surface roughness of the microcavity and thus the scattering losses, which predominantly arise due to cross-linking between the polymer layers, the toroid sensor is supplemented by a flexible support (hinge). It makes possible to fabricate a toroid sensor using the common layer-wise polymerization strategy and accommodate the demands for different orientations of the microresonator symmetry axis during the polymerization and sensing phases. Further scattering losses are determined by the cross-linking between the polymerization voxels and are minimized by a compromise between the voxel characteristics and focus spot translation speed. An exemplary image of the multiplexed sensor sample with 2PP toroid microcavities that are allocated at the relative offset of 100 $\mu$m is represented in Fig. 1.

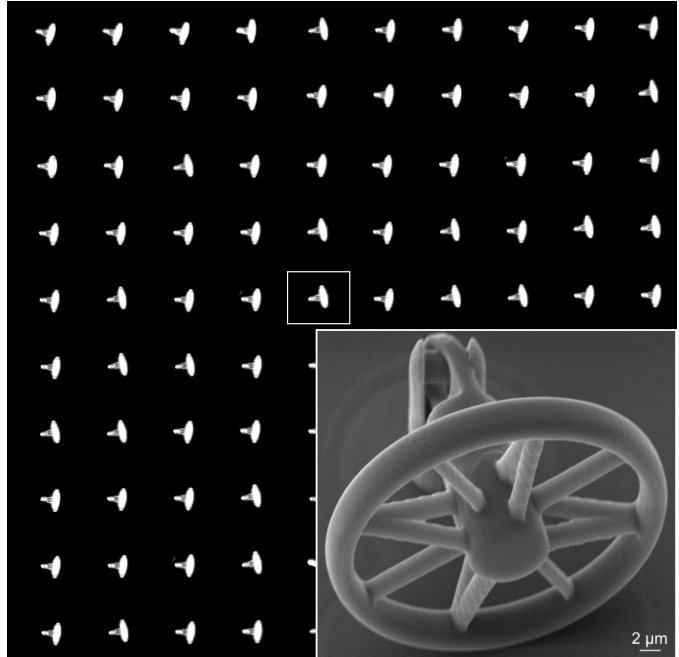

Figure 1: Overview of the multiplexed sensor with 100 microresonators taken by the camera of the measuring system. Inset demonstrates a SEM image of the individual microresonator.

**Sensing performance optimization** By optimizing the 2PP microresonator geometry and polymerization conditions, the intrinsic energy losses have been minimized. The remaining component of the loaded Q-factor – the coupling losses – describes the effectiveness of the energy transfer from the external medium to the microcavity. The coupling losses are minimized when the phase matching conditions and the overlap of the light field between the coupler and the cavity are met. By lining up the coupling frequency with the WGM and the propagation constants in the coupling and microcavity, phase matching is accomplished. For the optical prism coupling scheme this means equalizing the effective refractive index of the microresonator mode ($n_{eff}$) with in-plane prism refractive index ($n_c$) projection for the excitation beam: $n_{eff} = n_c \sin\theta$. The modes of the microtoroid have been numerically searched around 680 nm (refractive index of the coupling prism, N-BK7, is here 1.5136) within the coupling angles [37°:45°] in Comsol Multiphysics (Fig. 2a). Results show that the fundamental mode can be excited in the 2PP toroid microcavity with selected geometry ($R$ = 21 $\mu$m, $\rho$ = 1.8$\mu$m, $n_r$ = 1.502) at $\theta$ = 45°. By decreasing the angle, the efficient excitation of the higher-order modes is expected, where $Q_{rad}$ for higher order modes drops from $\sim 10^7$ ($\alpha_{in}$ = 45°) down to $\sim 10^5$ ($\alpha_{in}$ = 37°). The sensitivity on changes of the bulk refractive index (RIU) varies insignificantly, where for the fundamental TE mode it does not exceed 20 nm/RIU. Combination of the illumination angle of 45° with the right-angle prism simplifies the overall geometry of the illumination optical path and enables exceptional mechanical robustness.

An overlap between the evanescent fields could be adjusted by changing the distance between the coupler (prism) and the microresonator. The variations of the coupling-limited Q-factor have been numerically studied according to equation for the optical prism coupling scheme derived in [16]. The results represented in Fig. 2b show that without a gap between the coupling prism and microresonator the maximum possible loaded Q-factor does not exceed $10^3$. With gap increasing the losses drop so that at 200 nm distance the loaded Q-factor may exceed $10^4$, and at 400 nm reaches $\sim 10^6$. The coupling limited Q-factor approaches the fundamental limit for absorption at the distance of more than 550 nm.

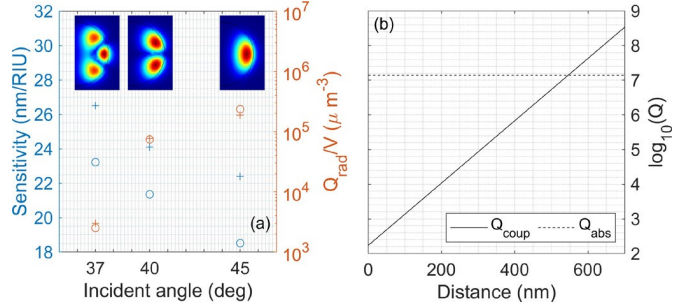

Figure 2: Numerical results on optimization of the coupling-limited Q-factor for the right-angle prism coupling scheme. (a) Impact of the incident angle $\alpha_{in}$ on sensitivity and $Q_{rad}/V$ ratio for the modes (circle - TE, cross - TM polarization) of different orders excited in the microcavity around 680 nm. Inserts demonstrate distribution of the EM field for TE modes phase matched at $\alpha_{in}$ = 37, 40 and 45 °. (b) Variations of the coupling-limited Q-factor ($Q_{coup}$, solid line) and absorption-limited Q-factor ($Q_{abs}$, dashed line) for different gaps between the coupling unit and microresonator.

For chosen sensor configuration, the gap may be ensured by the adhesive layer with the same refractive index as the intended external environment (water). However, the deposition of such layer is possible only before polymerization and, therefore, it would impact the photopolymerization process during the microcavities fabrication. The variations of the polymerization voxel caused by a thin distancing layer has been numerically studied using the model that describes the variations of the polymerization voxel dimensions when the 2PP illumination path lies through multiple layers with different optical properties [31].

A multilayer structure consisting of $m + 1 = 4$ layers: immersion oil (0), glass substrate (1), water-matched polymer layer (2) and photoresin (3) is simulated. The pulsed laser settings are: wavelength $\lambda$ = 780 nm, pulse duration $\tau$ = 90 fs, repetition frequency $f_{rep}$ = 82 MHz, average power $P_{avr}$ = 30 mW, laser spot translation speed equals to 100 mm/s, and polarization is linear, aligned along the X-axis. The 100× magnification objective lens has been selected with NA = 1.4, $f$ = 200 $\mu$m. Refractive index of the material for distancing layer set to $n_2$ = 1.328 - 6.2*10$^{-7}i$ (MyPolymer MY-133MC) and refractive index for other materials are $n_0$ = 1.518, $n_1$ = 1.5168, $n_3$ = 1.502. Thickness of the glass substrate is $th_1$ = 150 $\mu$m which is supposed to have a perfect plain surface ($\sigma_1 = 0$). The thickness of the distancing film has been varied within the value of the wavelength (from 0 to 700 nm). The calculation region has been limited by 0.8 $\mu$m for X and Y, and by 3 $\mu$m for Z coordinate with the grid size of $\lambda/100$.

It has been calculated that the light transmission level remains at $\approx$ 90% with reflection of up to 10% (absorption can be neglected) for the thickness of the water-matched layer within one wavelength. The essential portion of the energy that is reflected back enables enhancement in accuracy of the substrate/photoresin plane localization that is one of the most critical aspects limiting the 2PP manufacturing repeatability [20]. In turn, this avoids missing microresonators in the array due to the lack of contact with the substrate (washed out during development) or damage to the underside (submerged into the substrate). The variations of the characteristic length of the polymerization voxel (volume cube root) along the whole polymerization depth valid for chosen microresonator geometry (0-43 $\mu$m) are represented in Fig. 3.

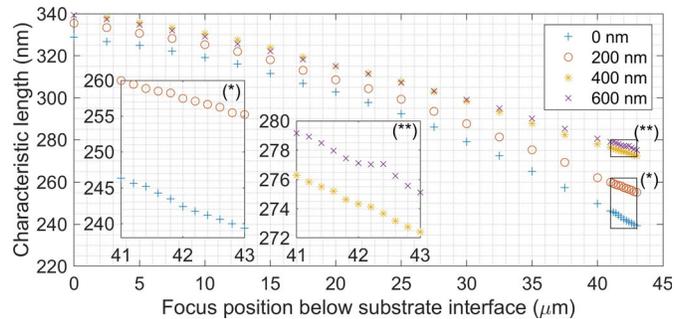

Figure 3: Results on numerical estimations of deviations of the polymerization voxel characteristic length for different thicknesses of the water-matched adhesive film (0, 200, 400, 600 nm) between the substrate and photosensitive material in the 2PP process.

The mean power of the laser is tuned for different thicknesses of the distancing layer in order to compensate the reflected energy and equalize the energy delivered to the photosensitive material. Particularly this means the increase of the laser mean power on 12.5%, 10%, and 4.5% relative to the case without distancing layer for 600, 400, and 200 nm thick film, correspondingly. The results show that the voxel length continuously declines with the distance to the substrate and the character of variations for the illumination configuration with water-matched layer is in accordance with the one without additional distancing layer. An impact of the thin adhesive film is determined to be insignificant, where the major voxel length deviations are linked to the refractive index mismatch between the glass substrate, immersion oil and photoresin material. At the same time, the variations of the characteristic voxel size remains limited by 5 nm around the region of [41:43] $\mu$m where the material is illuminated to form a microresonator for illumination configuration with adhesive film. According to the model that estimates scattering losses limited Q-factor ($Q_{scat}$) as a function of the varying relative permittivity proposed in [15] the determined level of variations of the voxel characteristics ensures $Q_{scat} \sim 10^6$. This convinces the ability to produce the high-quality polymer surfaces with optical quality even in the presence of a sub-wavelength distancing layer.

In order to form a homogeneous distancing layer with the thickness of several hundreds of nanometers, the viscosity of the original water-matched polymer (MY-133MC, MyPolymers) has been reduced by mixing with a solvent. Adhesive solutions at different proportions with the solvent starting from 1:15 up to 1:80 have been deposited onto the substrate and spin coated at 3000 rpm over 30 s and then the layers thickness have been studied with white light interferometer. It has been determined that 1:15 and 1:16 solutions lead to formation of the distancing layer with thickness above the laser's wavelength, whereas for 1:80 solution the layer was not detected. For 1:18 solution the mean thickness of the distancing layer is measured at $\approx$ 550 nm, 400 nm layer distancing layer is measured for 1:22 solution, 300 nm layer for 1:26, and 200 nm for 1:30 solution. It has been determined that by increasing the gap, the observed loaded Q-factor in the water can be increased by two orders of magnitude. For the distance of 400 nm, where the theoretical limits for scatting ($10^6$) and coupling losses match, the upper limit for the loaded Q-factor of 2PP resonators has been measured at the level of $10^5$ (Fig. 4). Despite the accuracy on the order of tens of nanometers for positioning and structuring when fabricated via 2PP, the toroid resonators still exhibit spectral deviations with the median value of $4\times 10^4$, where it drops down to $2\times 10^4$ for particular resonances. Further distancing of toroids from the substrate does not enhance the loaded Q-factors, but results in more microcavities without detectable resonance behavior with less than 5% of toroids showing a signal for 550 nm gap. Impossibility to reach the theoretical limit is expected to be caused by further scattering losses on imperfections of the toroidal rim and polymer inhomogeneities.

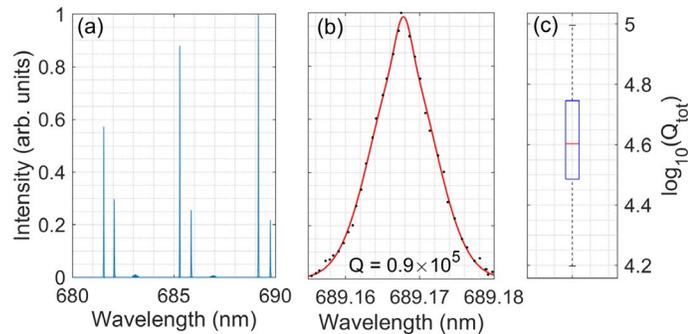

Figure 4: Spectral resonance properties of the exemplary 2PP toroid microresonator fabricated on a sample with 400 nm gap to the substrate. (a) Overview of the resonance lines in 10 nm spectral range. (b) Zoom into the resonance line with sharp peak. (c) Statistics on measured loaded quality factor in water environment for 100 microresonators on a single sample.

**4D sensing with 2PP resonators** The use of the 2PP microresonators fabricated out of the SZ2080 material is particularly attractive due to the self-sensing effect of the microresonator structure that enables realization of the 4D printed sensor system. The mechanism of self-sensing arises from swelling and shrinking forces that act on resonator geometry when immersed in different liquids. Particularly, when being immersed in the wetting solvents, the molecules of the latter start to penetrate into the polymer structure which leads to resonator expansion. When changing the environment to the non-wetting solvent, e.g., water, the penetrated molecules can be extracted and material shrinks. Here the overall gain for sensitivity relative to the numerical estimations for the pure evanescent field interaction ($\approx$ 20 nm/RIU, see Fig. 2) can rise on the orders of magnitude. Moreover, the efficiency of the self-sensing phenomenon tends to grow up when repeatedly inducing the sequential swelling and shrinking of microresonators, but after several repetitions of the swelling/shrinking actions the behavior stabilizes. Spectral response of the multiplexed 4D sensor (averaged over $23\times 23$ multiplexed sensor) for three repetitions in a row on the measurement of the phosphate buffered saline (PBS) solutions and water is represented in Fig. 5.

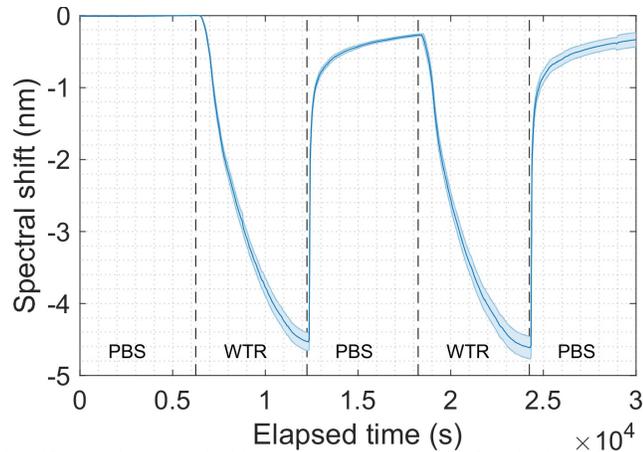

Figure 5: Evidence of the 4D sensing enabled by exemplary sample of multiplexed sensor with 529 units of 2PP microresonators in form of the long-term dynamics of the spectral shift for periodically varied environment between water (WTR) and phosphate buffered saline (PBS). Solid line shows spectral shift mean value, shaded area - 95% confidence interval.

The results show a clear distinction of the represented spectral shift variations against the expected step-like changes with short-term (∼300 s) dynamics for mixing the liquids in the sensing chamber in case of pure bulk refractive index sensing. Except for the first phase of PBS pumping with constant spectral response (sensor kept in PBS overnight before start of the measurement), the detected spectral shift variations show clear long-term dynamics for all phases, where the saturation is not gained within a phase. At the same time, the stability of the initial PBS state serves as a saturation level indicator for PBS measuring phases. The magnitude of the averaged spectral shift exceeds 4 nm over 6000 s of solution pumping that in comparison to ≈40 pm of spectral shift for bulk refractive index difference between PBS and water stands for swelling of the microresonators in PBS and their shrinkage in water. We have also observed the acceptable reproducibility of the mean spectral shift dynamics for water and PBS among repetitions, whereas the difference in sensing performance between microresonators leads to the increase of the confidence interval with time after beginning of the experiment. This is linked to the varying spectral properties of the individual 2PP microresonators.

**Perfluoroalkyl substances detection** At first, the microresonator response to PFAS chemicals is tested within the experiments where the environment in the sensing cell is sequentially changed from the clean deionized to PFOA contaminated water. Within these trials relatively high levels of contaminations (0.1 ppm-10 ppm) have been utilized. Unlike the averaged spectral shift, the spectral response (generalized) is retrieved hereinafter as the first principle component (PC1) which is a linear combination of the spectral shift dynamics among all resonators and follows the most prominent variance in the dataset. Such representation of the spectral variations is advantageous to account the different sensitivities of the microresonators and to reduce the impact of local variations/noises that is classified via principle component analysis (PCA) as insignificant component. It has been determined that the generalized spectral shift of the microresonator remains implicit with respect to the pure water for concentration up to 10 ppm. When changing water to 10 ppm PFOA the generalized response shows a clear negative trend that indicates microresonator shrinkage induced by PFOA molecules (Fig. 6).

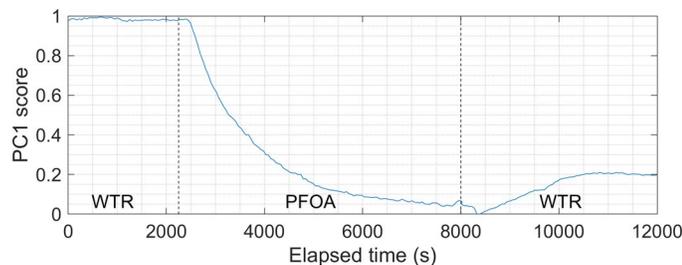

Figure 6: Spectral response of the multiplexed 4D microresonator sensor for 10 ppm PFOA solution followed by the pure water (WTR) flushing.

The measured spectral shift negative dynamics tends to reach a constant level at ≈6000 s after starting the pumping of the PFOA contaminated water. When the sensing chamber is refilled with clean water next to PFOA, the level of the generalized spectral response rises, but it does not recover to the initial water level. This is assumed to be related to the compression limit for the microresonator and the common character of the material response for pure water and PFOA (it tends to shrink). Therefore, the microcavity pre-swelling is expected to improve PFAS sensitivity and this can be done via pumping the PBS solutions (see Fig. 5). For this reason, the experiment has been designed so that the PBS is pumped for 6000 s before the PFAS sensing and prior the first phase of PFAS sensing the pure water was pumped for the same duration to get the identical state of the 4D microresonator.

The results on spectral variations when incubating multiplexed 4D sensor in increasing concentrations of PFOA and PFBS solutions in the range from 1 ppb to 20 ppm is represented in Fig. 7. Since both pure and contaminated water will initiate cavity shrinkage, the characteristic dynamical changes measured for PFAS-containing solution has been compared to the one for pure water. In order to convince that the sensing response of the microresonator does not alter with multiple runs of PFAS solution pumping, the water response has been measured before (WTR1) and after multiple interactions with contaminated water (WTR2). In order to enable comparison of PFAS-initiated dynamics that has been measured in the different experimental runs, the generalized spectral shift has been scaled.

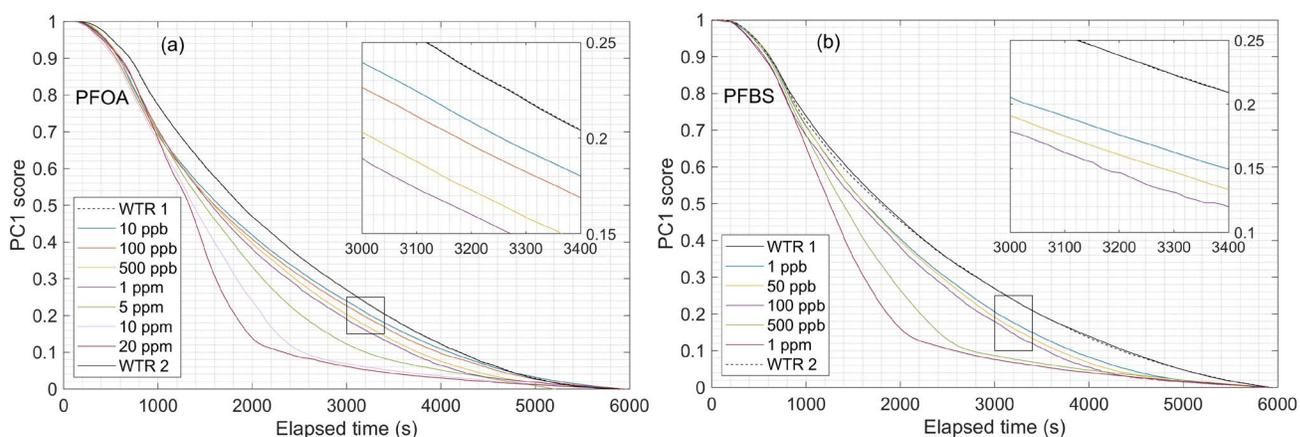

Figure 7: Dynamical variations of the generalized sensor response in form of the scaled PC1 for different concentrations of continuously pumped water-based solutions containing different PFAS substances at varied concentrations. Each variation is measured after PBS pumping phase for 6000 s. Spectral shift dynamics in pure water before (WTR1) and after (WTR2) measurements of PFAS concentrations are added to each curve. (a) Reaction on PFOA solutions with concentrations from 10 ppb to 20 ppm. (b) Reaction on PFBS solutions with concentrations from 1 ppb to 1 ppm.

We determined that presence of a small amount of PFOA or PFBS in water at the level of 10 ppb and 1 ppb correspondingly leads to changes in dynamics of the spectral changes relative to the uncontaminated water. The more PFAS components are added to the water, the higher is negative slope in dynamics that indicates an acceleration of the shrinkage process in presence of PFAS chemicals. Unlike FPBS, the difference between the pure water and PFOA-containing water is clearly seen already at first stages of the shrinkage process, but the difference in dynamics for various levels of PFBS contamination shows up only after hundreds of seconds of pumping the solutions through the sensor. In addition to the rise of the negative slope, the shape of the trend changes with PFAS concentration. For large concentrations (at the level of ppm), we have observed the appearance of two phase trend with acceleration in dynamics of the sensor response in the first phase and deceleration in the second phase in comparison to the one for water. Here for the first phase the difference between concentrations is more expressed than for the second phase and the transition point between the phases occurs earlier when increasing the concentration. For PFBS the transition point for 0.5 ppm appears at ≈2600 s, whereas for 1 ppm already at ≈2100 s. In general, the sensor is more sensitive to PFBS than to PFOA contaminations, where the sensor responses are comparable between 10 ppm PFOA and 0.5 ppm PFBS as well as between 20 ppm PFOA and 1 ppm PFBS, respectively.

The measured spectral response of 4D microresonators to the presence of PFAS molecules can primarily be attributed to the synergistic effect, which occurs when several factors for interaction of PFAS substances with polymer material act together. Due to the presence of the radical quencher (DMAEMA) in the two-photon polymerization process, which becomes a part of the polymer chain, the resulting microresonators contain amine functional groups. Materials with amine functional groups have shown performance in coupling with the anionic PFAS (which include the chemicals being tested) [1]. First of all, the interaction of the amine-containing material and PFAS molecule has electrostatic and hydrophobic nature, but microresonator morphology and porosity also impact the sensor response. The amines of the polymer and the carboxylate or sulfonate headgroup of the anionic PFAS are expected to interact electrostatically, where the pH value of the solution strongly impacts the process (the lower pH, the stronger interaction). The hydrophobic interactions are determined by the type of the head functional group, PFAS chain length, and hydrophobicity of the microresonator material. It has been reported that PFAS chemicals with sulfonic headgroups (PFBS) have a better adsorption capability than the ones with carboxylic headgroups (PFOA) of the same carbon number [9]. In another work [14], where PFAS chemicals of different chain length have been compared, it was revealed that the adsorption of the longer-chain compounds is always higher than their shorter-chain counterparts. Two explanations exist for this: the first points that the long-chain PFAS interact both electrostatically and hydrophobically while short-chain ones primarily electrostatically; another addresses the increased potential of multilayer formation for long-chain PFAS. Morphology and porosity of the material are crucial for PFAS diffusion dynamics and accessibility of the adsorption sites, where lower material surface area and/or long-chain chemicals result in slower kinetics [42]. Particularly increased efficiency in sensing of the PFBS components in comparison to PFOA is expected to be linked with the length of the chain. Being longer-chain compound PFOA is expected to be less capable of penetrating into the nanopores of the microresonator and thus the speed of the material shrinkage is less than for PFBS. For high PFAS concentrations the role of the small variance in bulk refractive index of solutions, that is related to the chemical composition of the perfluorinated compounds, rises [5]. This is expressed by the appearance of the two-phases dynamics for higher concentrations of the PFAS components which is expected to be related to a slight decrease in the bulk refractive index of the solutions. This governs the difference in dynamics between the same concentrations of PFOA and PFBS (in the range of ppm), where the latter shows faster resonator shrinkage.

## Conclusions

In this paper we reported on detection of the per- and polyflouroalkyl water contaminants with optical microresonator sensor to the best of our knowledge demonstrated for the first time. The sensor has been constructed in the multiplexed imaging configuration with multiple hundreds of 4D microcavities fabricated with two-photon polymerization that enable enhanced sensitivity thanks to the material reaction on the external medium (self-sensing). The paper presents a list of numerical and experimental improvements implemented in the fabrication of an array of 2PP microresonators that improve spectral response by up to two orders of magnitude for 4D sensors compared to their 3D counterparts and demonstrate a loaded quality factor in water of $\sim 10^5$. The performance in detection of PFAS contaminations in water has been tested on example of two representatives of the class of anionic substances with different functional groups and chain lengths (PFOA and PFBS). We demonstrated that concentrations of 1 ppb PFBS and 10 ppb PFOA are detectable without applying selective layers onto microresonators. Given the key advantages of the microresonator-based optical sensors, the applicability of this approach is expected to boost the registration of the persistent water contaminants. With further studies on the host material for the 4D microresonator unit, it is expected to be potentially feasible to implement an integrated approach for solving the problem of contamination of the aqueous environments with PFAS substances with simultaneous detection of the initial contamination and efficient removal of small PFAS concentrations.

## Materials and methods

**Chemicals** Sol-gel SZ2080 photoresin combined with 4,4'-Bis-(diethylamino)-benzophenon photoinitiator and another monomer, 2-(dimethylamino) ethyl methacrylate (DMAEMA), provided by the FORTH Center has been selected for fabrication of the microcavity sensors [33]. The second monomer in this composition serves as a radical quencher throughout the polymerization process in order to realize the diffusion-assisted 2PP. Phosphate buffer saline (PBS) has been prepared in the deionized water yielding 0.01 M phosphate buffer, 0.0027 M potassium chloride and 0.137 M sodium chloride, pH 7.4, at 25 °C. Perfluorooctanoic acid (PFOA), perfluorobutanesulfonic acid (PFBS) of analytical grade were purchased from Sigma-Aldrich and dissolved in the deionised water in concentrations ranging from ppb (ng/ml) to ppm ($\mu$g/ml).

**Sensor fabrication** Cover glasses with thickness of 150 $\mu$m have been selected as the substrates for fabrication of the arrays of polymer toroid microresonators. Prior to casting a drop of the photoresin solution a 400 nm layer of the water-matched refractive index adhesive has been deposited onto the substrate via spin-coating at 3000 rpm for a duration of 30 s. The substrate was heated at 50 °C for 4 hours to evaporate the solvent and guarantee an effective contact between the polymer structure and the substrate. The substrate was then cooled down to the room temperature and then placed into the 2PP setup.

Two-photon polymerization fabrication is performed on the basis of the in-house made setup with a mode locked Ti:Sa laser system (Tsunami, Spectra Physics) with emission wavelength of 780 nm, repetition rate of 82 MHz, and the pulse width of 90 fs. Power of the incoming light is controlled by a photodiode and adjusted via combination of rotatable $\lambda/2$ retardation plate with a polarizing beamsplitter. Flexibility of the polymerization process in 3D space is guaranteed by an acousto-optical modulator (AOM) used as a shutter, galvo scanner that allows laser beam redirection in the plane parallel to the substrate, and positioning of the sample in height is enabled by the linear stages. Using an oil-immersed 100× objective with NA of 1.4, the laser is tightly focused into the photoresin to gain high resolution in fabrication of 3D polymer structures. To prevent mixing of the photosensitive material with immersion oil, the sample with photoresin droplet is placed upside-down. The computer-aided design (CAD) model of the microresonator is divided into layers and then the program creates a unique route for the laser beam for each layer.

We have optimized the set of illumination parameters to ensure the polymerization of a thousand toroid sensors in a reasonable amount of time. Slicing and hatching distances have been selected as 200 nm and 100 nm correspondingly. It has been discovered that the speeding-up of the laser spot translation up to 100 mm/s for toroid and up to 200 mm/s for the supporting components with corresponding enhancement of the delivered average laser power till $P_{avr}$ = 30 mW does not lead to formation of the structural combs on the cavity rim which would indicate the insufficient voxel overlapping. With this set of parameters, the duration of manufacturing of a single microresonator reduces to less than one minute. The 2PP samples were wet-chemically treated for 20 min in the 4-methylpentan-2-one developer before being submerged for 10 min in 2-propanol to remove the non-polymerized material. The constructions have finally been left in ambient conditions for a number of hours to allow the solvent to dissipate.

**Measuring instrument** Sensing with multiple polymer microresonators is based on collimated laser light illumination of the whole area containing cavities, excitation via an optical prism, and radiated light detection in the far field with a camera (Fig. 8).

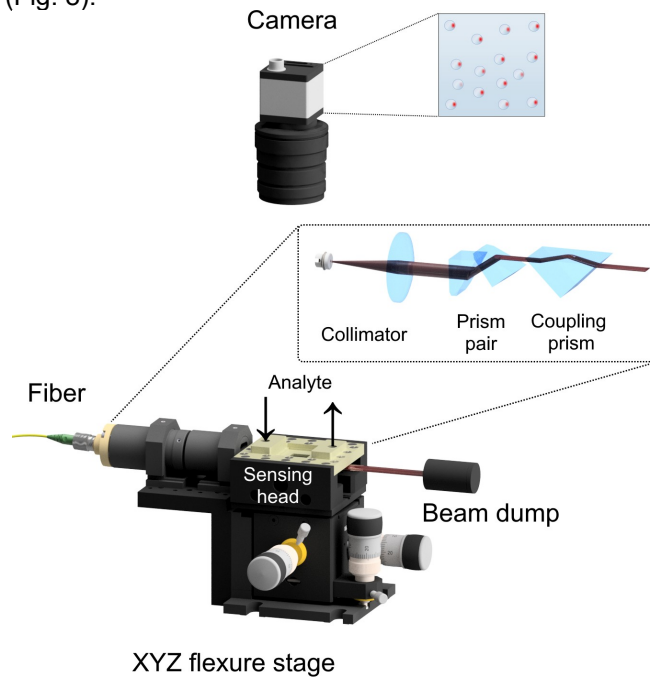

Figure 8: Main components of the sensing instrument for simultaneous collection of spectral responses from multiple polymer microresonator. Excitation laser, wavemeter, and components for fluid selection and pumping are not depicted.

Sensor excitation part is based on a diode laser (Velocity, New Focus) that is tunable from 680 to 690 nm with 200 kHz linewidth. Through the use of suitable single mode optical fibers (630HP, Thorlabs) and individual polarization controllers (FPC030, Thorlabs), the laser beam is transmitted to the sensor head to excite the TE modes. Constant feedback loop to the wavemeter (WS7-30, HighFinesse) with fiber splitters (99%/1%) is introduced in order to control the laser wavelength. The achromatic optical collimation package (60FC-T-4-M40-24, Schaefter+Kirchhoff) with an output beam diameter of ≈ 8 mm is used to achieve the illumination of the whole sensing area on the sample. Anamorphic prism pair is introduced into the optical path after the collimator to change the incident laser beam profile from circular to elliptical. This compensates the elongation of the collimated beam projection in the propagation direction at the prism excitation surface and ensures a circular shape of the spot on the prism excitation surface. Sensor detection part contains lens objective and a monochrome high-speed global shutter camera (CB262RG-GP-X8G3, Ximea).

For stabilizing the coupling conditions, an immersion oil is used to obtain optical contact between the cover glass of the sensor and the optical prism. The sensing head is put together using the prism holder, optical prism, sensor sample, and flow chamber components and positioned on a precise three-axis flexure stage. Microfluidic tubing connects the sensing head with the fluid pumping system from one side and with waste container from the other side. The pressure-based controller (LINEUP FLOW EZ, Fluigent) is used in conjunction with the flow rate sensor (FLOW UNIT, Fluigent) and selection valve (M-SWITCH, Fluigent) to sequentially pump fluids from 8 separate containers through the array of toroids at a regulated speed of 100 $\mu$l/min. Ambient temperature variations are measured with the temperature sensor (PT 100) embedded into the sensing head.

## Acknowledgement

We would like to acknowledge the group of Dr. Maria Farsari (IESL-FORTH) and particularly Dr. Gordon Zyla for providing the photoresin. The authors Andreas Ostendorf and Anton Saetchnikov are grateful to the German Federal Ministry for Research and Education (BMBF) for partly funding this work under the VIP+-Programme in the project IntellOSS, 03VP08220.

## Conflict of interest

The authors declare no conflict of interest.

## References

[1] Mohamed Ateia, Alaaeddin Alsbaiee, Tanju Karanfil, and William Dichtel. Efficient PFAS removal by amine-functionalized sorbents: Critical review of the current literature. *Environmental Science & Technology Letters*, 6(12):688–695, 2019.

[2] Mohamed Ateia, Amith Maroli, Nishanth Tharayil, and Tanju Karanfil. The overlooked short- and ultrashort-chain poly- and perfluorinated substances: A review. *Chemosphere*, 220:866–882, 2019.

[3] V. B. Braginsky, M. L. Gorodetsky, and V. S. Ilchenko. Quality-factor and nonlinear properties of optical whispering-gallery modes. *Physics Letters A*, 137(7-8):393–397, 1989.

[4] Lu Cai, Junyao Pan, Yong Zhao, Jin Wang, and Shiyuan Xiao. Whispering gallery mode optical microresonators: Structures and sensing applications. *physica status solidi (a)*, 217(6):1900825, 2020.

[5] Nunzio Cennamo, Girolamo D'Agostino, Gianni Porto, Adriano Biasiolo, Chiara Perri, Francesco Arcadio, and Luigi Zeni. A molecularly imprinted polymer on a plasmonic plastic optical fiber to detect perfluorinated compounds in water. *Sensors (Basel, Switzerland)*, 18(6), 2018.

[6] Nunzio Cennamo, Luigi Zeni, Paolo Tortora, Maria Elena Regonesi, Alessandro Giusti, Maria Staiano, Sabato D'Auria, and Antonio Varriale. A high sensitivity biosensor to detect the presence of perfluorinated compounds in environment. *Talanta*, 178:955–961, 2018.

[7] Weijian Chen, Şahin Kaya Özdemir, Guangming Zhao, Jan Wiersig, and Lan Yang. Exceptional points enhance sensing in an optical microcavity. *Nature*, 548(7666):192–196, 2017.

[8] Yongpeng Chen, Yin Yin, Libo Ma, and Oliver G. Schmidt. Recent progress on optoplasmonic whispering–gallery–mode microcavities. *Advanced Optical Materials*, 9(12):2100143, 2021.

[9] Ziwen Du, Shubo Deng, Yue Bei, Qian Huang, Bin Wang, Jun Huang, and Gang Yu. Adsorption behavior and mechanism of perfluorinated compounds on various adsorbents–a review. *Journal of hazardous materials*, 274:443–454, 2014.

[10] M. Eryürek, Z. Tasdemir, Y. Karadag, S. Anand, N. Kilinc, B. E. Alaca, and A. Kiraz. Integrated humidity sensor based on SU-8 polymer microdisk microresonator. *Sensors and Actuators B: Chemical*, 242(10):1115–1120, 2017.

[11] Fairuza Faiz, Gregory Baxter, Stephen Collins, Fotios Sidiroglou, and Marlene Cran. Polyvinylidene fluoride coated optical fibre for detecting perfluorinated chemicals. *Sensors and Actuators B: Chemical*, 312:128006, 2020.

[12] Heidelore Fiedler, Todd Kennedy, and Barbara J. Henry. A critical review of a recommended analytical and classification approach for organic fluorinated compounds with an emphasis on per- and polyfluoroalkyl substances. *Integrated environmental assessment and management*, 17(2):331–351, 2021.

[13] Matthew R. Foreman, Jon D. Swaim, and Frank Vollmer. Whispering gallery mode sensors. *Advances in optics and photonics*, 7(2):168–240, 2015.

[14] Erica Gagliano, Massimiliano Sgroi, Pietro P. Falciglia, Federico G. A. Vagliasindi, and Paolo Roccaro. Removal of poly- and perfluoroalkyl substances (PFAS) from water by adsorption: Role of PFAS chain length, effect of organic matter and challenges in adsorbent regeneration. *Water research*, 171:115381, 2020.

[15] Michael L. Gorodetsky, Andrew D. Pryamikov, and Vladimir S. Ilchenko. Rayleigh scattering in high-Q microspheres. *Journal of the Optical Society of America B*, 17(6):1051–1057, 2000.

[16] M. L. Gorodetsky and V. S. Ilchenko. Optical microsphere resonators: Optimal coupling to high-Q whispering-gallery modes. *Journal of the Optical Society of America B*, 16(1):147, 1999.

[17] Xuefeng Jiang, Abraham J. Qavi, Steven H. Huang, and Lan Yang. Whispering-gallery sensors. *Matter*, 3(2):371–392, 2020.

[18] Cédric Lemieux-Leduc, Régis Guertin, Marc-Antoine Bianki, and Yves-Alain Peter. All-polymer whispering gallery

mode resonators for gas sensing. *Optics express*, 29(6):8685–8697, 2021.

[19] Jie Liao and Lan Yang. Optical whispering-gallery mode barcodes for high-precision and wide-range temperature measurements. *Light: Science & Applications*, 10(1):32, 2021.

[20] Mangirdas Malinauskas, Maria Farsari, Algis Piskarskas, and Saulius Juodkazis. Ultrafast laser nanostructuring of photopolymers: a decade of advances. *Physics Reports*, 533(1):1–31, 2013.

[21] Giulia Moro, Francesco Chiavaioli, Stefano Liberi, Pablo Zubiate, Ignacio Del Villar, Alessandro Angelini, Karolien de Wael, Francesco Baldini, Ligia Maria Moretto, and Ambra Giannetti. Nanocoated fiber label-free biosensing for perfluorooctanoic acid detection by lossy mode resonance. *Results in Optics*, 5(2):100123, 2021.

[22] Shoji F. Nakayama, Mitsuha Yoshikane, Yu Onoda, Yukiko Nishihama, Miyuki Iwai-Shimada, Mai Takagi, Yayoi Kobayashi, and Tomohiko Isobe. Worldwide trends in tracing poly- and perfluoroalkyl substances (PFAS) in the environment. *TrAC Trends in Analytical Chemistry*, 121(10):115410, 2019.

[23] A. Ovsianikov, A. Ostendorf, and B. N. Chichkov. Three-dimensional photofabrication with femtosecond lasers for applications in photonics and biomedicine. *Applied Surface Science*, 253(15):6599–6602, 2007.

[24] Bilal Özel, Ralf Nett, Thomas Weigel, Gustav Schweiger, and Andreas Ostendorf. Temperature sensing by using whispering gallery modes with hollow core fibers. *Measurement Science and Technology*, 21(9):094015–094020, 2010.

[25] Rosalba Pitruzzella, Francesco Arcadio, Chiara Perri, Domenico Del Prete, Giovanni Porto, Luigi Zeni, and Nunzio Cennamo. Ultra-low detection of perfluorooctanoic acid using a novel plasmonic sensing approach combined with molecularly imprinted polymers. *Chemosensors*, 11(4):211, 2023.

[26] Tess Reynolds, Nicolas Riesen, Al Meldrum, Xudong Fan, Jonathan M. M. Hall, Tanya M. Monro, and Alexandre François. Fluorescent and lasing whispering gallery mode microresonators for sensing applications. *Laser & Photonics Reviews*, 11(2):1600265, 2017.

[27] Heejeong Ryu, Baikun Li, Sylvain de Guise, Jeffrey McCutcheon, and Yu Lei. Recent progress in the detection of emerging contaminants PFASs. *Journal of hazardous materials*, 408:124437, 2021.

[28] Anton V. Saetchnikov, Elina A. Tcherniavskaia, Vladimir A. Saetchnikov, and Andreas Ostendorf. Intelligent optical microresonator imaging sensor for early stage classification of dynamical variations. *Advanced Photonics Research*, 7:2100242, 2021.

[29] Anton V. Saetchnikov, Elina A. Tcherniavskaia, Vladimir A. Saetchnikov, and Andreas Ostendorf. A laser written 4D optical microcavity for advanced biochemical sensing in aqueous environment. *Journal of Lightwave Technology*, 38(8):2530–2538, 2020.

[30] Anton V. Saetchnikov, Elina A. Tcherniavskaia, Victor V. Skakun, Vladimir A. Saetchnikov, and Andreas Ostendorf. Reusable dispersed resonators-based biochemical sensor for parallel probing. *IEEE Sensors Journal*, 19(17):7644–7651, 2019.

[31] Anton Saetchnikov, Vladimir Saetchnikov, Elina Tcherniavskaia, and Andreas Ostendorf. Effect of a thin reflective film between substrate and photoresin on two-photon polymerization. *Additive Manufacturing*, 24:658–666, 2018.

[32] Anton Saetchnikov, Elina Tcherniavskaia, Vladimir Saetchnikov, and Andreas Ostendorf. Deep-learning powered whispering gallery mode sensor based on multiplexed imaging at fixed frequency. *Opto-Electronic Advances*, 3(11):200048, 2020.

[33] Ioanna Sakellari, Elmina Kabouraki, David Gray, Vytautas Purlys, Costas Fotakis, Alexander Pikulin, Nikita Bityurin, Maria Vamvakaki, and Maria Farsari. Diffusion-assisted high-resolution direct femtosecond laser writing. *ACS nano*, 6(3):2302–2311, 2012.

[34] Brian F. Scott, Cheryl A. Moody, Christine Spencer, Jeffrey M. Small, Derek C. G. Muir, and Scott A. Mabury. Analysis for perfluorocarboxylic acids/anions in surface waters and precipitation using GC–MS and analysis of PFOA from large-volume samples. *Environmental science & technology*, 40(20):6405–6410, 2006.

[35] Mariana P. Serrano, Sivaraman Subramanian, Catalina von Bilderling, Matías Rafti, and Frank Vollmer. "Grafting-To" covalent binding of plasmonic nanoparticles onto silica WGM microresonators: Mechanically robust single-molecule sensors and determination of activation energies from single-particle events. *Sensors (Basel, Switzerland)*, 23(7), 2023.

[36] Judith Su, Alexander Fg Goldberg, and Brian M. Stoltz. Label-free detection of single nanoparticles and biological molecules using microtoroid optical resonators. *Light: Science & Applications*, 5(1):e16001, 2016.

[37] Nikita Toropov, Gema Cabello, Mariana P. Serrano, Rithvik R. Gutha, Matías Rafti, and Frank Vollmer. Review of biosensing with whispering-gallery mode lasers. *Light: Science & Applications*, 10(1):42, 2021.

[38] Kerry J. Vahala. Optical microcavities. *Nature*, 424(6950):839–846, 2003.

[39] F. Vollmer, D. Braun, A. Libchaber, M. Khoshsima, I. Teraoka, and S. Arnold. Protein detection by optical shift of a resonant microcavity. *Applied Physics Letters*, 80(21):4057–4059, 2002.

[40] Yi Xu, Ping Bai, Xiaodong Zhou, Yuriy Akimov, Ching Eng Png, Lay-Kee Ang, Wolfgang Knoll, and Lin Wu. Optical refractive index sensors with plasmonic and photonic structures: Promising and inconvenient truth. *Advanced Optical Materials*, 7(9):1801433, 2019.

[41] Deshui Yu, Matjaž Humar, Krista Meserve, Ryan C. Bailey, Síle Nic Chormaic, and Frank Vollmer. Whispering-gallery-mode sensors for biological and physical sensing. *Nature Reviews Methods Primers*, 1(1):453, 2021.

[42] Qiang Yu, Ruiqi Zhang, Shubo Deng, Jun Huang, and Gang Yu. Sorption of perfluorooctane sulfonate and perfluorooctanoate on activated carbons and resin: Kinetic and isotherm study. *Water research*, 43(4):1150–1158, 2009.